\documentclass[a4paper,11pt]{article}
\pdfoutput=1 

\usepackage{jheppub} 

\usepackage{multirow}
\newcommand{\ba}{\begin{eqnarray}}
\newcommand{\ea}{\end{eqnarray}}
\usepackage{textcomp}
\usepackage{amsmath}

\usepackage[T1]{fontenc} 

\title{\boldmath Charged traversable wormholes supported by Casimir energy with and without GUP corrections}

\author[a]{Daris Samart,}
\author[b]{Takol Tangphati,}
\author[c,d]{Phongpichit Channuie}

\affiliation[a]{Khon Kaen Particle Physics and Cosmology Theory Group (KKPaCT), Department of Physics, Faculty of Science, Khon Kaen University, 123 Mitraphap road, Khon Kaen, 40002, Thailand}
\affiliation[b]{Theoretical and Computational Physics Group, Theoretical and Computational Science Center (TaCS), Faculty of Science, King Mongkut’s University of Technology Thonburi, Prachauthid Rd., Bangkok 10140, Thailand}
\affiliation[c]{School of Science, Walailak University, Thasala, Nakhon Si Thammarat, 80160, Thailand}
\affiliation[d]{College of Graduate Studies, Walailak University, Thasala, Nakhon Si Thammarat, \\80160, Thailand}

\emailAdd{darisa@kku.ac.th,takoltang@gmail.com,channuie@gmail.com}

\abstract{In this paper, we investigate new exact and
analytic solutions of the Einstein–Maxwell field equations
describing Casimir wormholes with and without the effect of the Generalized Uncertainty Principle (GUP). We consider a specific type of the GUP relations and study three specific models of the redshift function along with two different EoS of state given by $p_{r}(r)=\omega_{r}(r)\rho(r)$ and $p_{t}(r)=\omega_{t}(r)p_{r}(r)$. Here we obtain a class of asymptotically flat Casimir wormhole solutions with and without GUP corrections under the effect of electric charge. Furthermore we check the null, weak, strong and dominant energy conditions at the wormhole throat of radius $r_{0}$, and show at the wormhole throat that the classical energy conditions are violated by some arbitrary small quantity. We also examine the wormhole geometry with semi-classical corrections by using embedding diagrams. We use the volume integral quantifier to quantify the amount of the exotic matter near the wormhole throat. Additionally, we investigate exotic fluid near WH throat using the so-called exoticity parameter and discuss the speed of sound.}
\begin{document} 
\maketitle
\flushbottom

\section{Introduction}
Wormholes are hypothetical objects that are used to connect different universes or different spatial points in the same universe \cite{Visser1995}. The first theoretical investigation of the existences of the wormholes received significant attention after the discovery of GR shortly by Flamm \cite{Flamm:1916}. Subsequently, a detailed study of the wormholes has been further extended by Einstein and Rosen \cite{Einstein:1935tc} and a so-called Einstein-Rosen bridge was used to represent wormholes at that moment. However, it was realized later that the Einstein-Rosen bridge can not be traversable. Interestingly, the effect of the electromagnetic interaction in Einstein field equation might play a crucial role on creating wormholes by using geon \cite{Wheeler:1955zz,Misner:1957mt}. However, early studies of the wormholes are just thought experiments or theoretical playgrounds which are unable to contact to physical realities. Until, Morris and Thorne first studied the possible existence and demonstrated how to use the wormholes for interstellar travel \cite{Morris:1988cz}. Nevertheless, a so-called exotic matter with negative pressure is required for supporting the throat of traversable wormholes. This leads to violations of several (classical) energy conditions in GR. Therefore, searching for the forms of exotic matter compatible with the energy conditions becomes the main topics in this research field.

A number of exotic matter forms has been studying in the traversable wormholes. On the one hand, modified theories of gravity are used to generate effective exotic fluid in order to support the throat of traversable wormholes. For instances, these include higher order gravity and its motivations by string theory \cite{Giribet:2019dmg,Ovgun:2017dik,KordZangeneh:2015dks,Mehdizadeh:2012zz,Dehghani:2011fa,Dehghani:2009zza,Rahaman:2006xb,Kim:1996np,Ghoroku:1992tz,MenaMarugan:1991ea,Hochberg:1990is}, alternative theories of quantum gravity, e.g., Horava-Lifshitz 
\cite{Garcia-Compean:2020aaa,Bellorin:2016nvh,Bellorin:2015oja,Bellorin:2014qca,Botta-Cantcheff:2009ffi} and massive gravity \cite{Kamma:2021wam,Amirabi:2020kfr,Tangphati:2019pxh,Forghani:2018svt,Paul:2018ppy,Jusufi:2017drg}, $f(R)$ gravity \cite{Mishra:2021ato,Eid:2020hvf,Shamir:2020uzy,Fayyaz:2020jzh,Tangphati:2020mir,Godani:2019kgy,Samanta:2019tjb,Godani:2018blx,Sharif:2018jdj,Saeidi:2011zz,Bronnikov:2010tt,Lobo:2009ip}, scalar-tensor (Brans-Dicke) gravity and its generalizations (both Horndeski and Galileon theories)
\cite{Korolev:2020ohi,Papantonopoulos:2019ugr,Franciolini:2018aad,Mironov:2018uou,Mironov:2018pjk,Evseev:2017jek,Rubakov:2016zah,Kolevatov:2016ppi,Bhattacharya:2009rt,Eiroa:2008hv,Lu:2003bg,He:1999nd,Nandi:1997en,Agnese:1995kd,Xiao:1991nv}, Einstein-Gauss-Bonnet thoery 
\cite{Sharif:2020xxj,Ibadov:2020btp,Antoniou:2019awm,Mehdizadeh:2015jra,Kanti:2011yv,Kanti:2011jz,Mazharimousavi:2010bm,Maeda:2008nz,Bhawal:1992sz}, teleparallel gravity 
\cite{Singh:2020rai,Mustafa:2021vqz,Saaidi:2020rrb,Bahamonde:2016jqq,Jawad:2015uea,Salti:2013bha,Boehmer:2012uyw,Aygun:2007zzb}, and etc. On the other hand, exotic matter motivated from dark energy with standard Einstein-Hilbert action also receive a lot of attentions, e.g., vacuum energy and cosmological constant 
\cite{Santos:2021jjs,Ambjorn:2021wdm,Jusufi:2020rpw,Garattini:2019ivd,Heydarzade:2014ada,Rahaman:2006xa,Lemos:2003jb,Liu:1993ej,Nishioka:1992xy,Myers:1990mjh,Klebanov:1988eh}, quintessence and phantom scalar fields \cite{Aounallah:2020rlf,Zubair:2020uyb,Manna:2019tpn,Dzhunushaliev:2017syc,Kocuper:2017aap,Dzhunushaliev:2010bv,Cataldo:2008ku,Lobo:2006cx,Rahaman:2005qi,Zaslavskii:2005fs,Popov:2001kk,Popov:2000id,Kim:1992wd}, vector gauge field and its higher-order forms \cite{Anand:2020wlk,Barros:2018lca,Cox:2015pga,Rahaman:2006qm,Tan:1997zc,Yoshida:1990quk,Shen:1990cb}, and Chaplygin gas fluids 
\cite{Lobo:2005vc,Jamil:2008wu,Eiroa:2009hm,Kuhfittig:2009mx,Sharif:2013lna,Sharif:2013tva,Elizalde:2018frj}. In result, both modified gravity theories and inclusion of special forms of exotic matter could stabilize the wormhole throat. Although most of the models are still very toy models that are not capable of being tested in the laboratories as well as suffering from some theoretical inconsistencies in the models themselves. However, a so-called Casimir energy is one type of the vacuum energy which has been confirmed by the experiment. Moreover, the Casimir energy is the most likely candidate of matter forms that can be used to stabilize the traversable wormholes. In addition, it is well known that the classical electric and magnetic fields could be useful to assist the stabilization of the wormholes. Possibility of wormholes having charge was proposed by Ref.\cite{Kim:2001ri}. The charges play the role of the additional matter to the static wormhole which is stabilized by the exotic matter. Moreover, wormholes with charge are thought to be a natural extension of the original Morris-Thorne wormhole \cite{Kuhfittig:2011xh}. 

Then the main purpose of this work is to study the physically possible construction of the Casimir wormholes with the electric charge underlying a so-called Generalized Uncertainty Principle (GUP). In this work, we would like to extend the analysis present in Ref. \cite{Jusufi:2020rpw} where the effect of GUP have been investigated in the Casimir wormholes. In addition,  Ref.\cite{Boyer:1968uf} have studied the physical implications of the Casimir effect on a charge particle in a conducting spherical shell. Therefore, inclusion of the electric charge in the GUP Casimir wormholes is worth studying in more details in order to see how the electric field plays the role in wormhole constructions both with and without the GUP by invoking several models of the red shift functions. 

This work is organized as follows: We consider the charged Casimir wormholes with and without the GUP corrections in Sec.\ref{ccwh}. Subsequently, the embedding diagrams of the wormholes are illustrated in Sec.\ref{embed}. In Sec.\ref{ec}, we study the energy conditions of the charge Casimir wormholes. The amount of the exotic matter near the wormhole throat will be calculated in Sec.\ref{exotic-matter}. Additionally, in Sec.\ref{sec6} we investigate exotic fluid near WH throat using the so-called exoticity parameter and discuss the speed of sound. We close this work in Sec.\ref{conclude} for our conclusion of findings and discussion relevant physical consequences and future outlook.

\section{Casimir wormhole spacetime with electric charge}
\label{ccwh}
The existence of charged black holes has suggested that wormholes may also be charged. A generic static and spherically symmetric WH can be described by the Morris-Thorne (MT) metric \cite{Morris:1988cz}. Here we begin with MT wormhole spacetime in the Schwarzschild coordinates given by
\begin{eqnarray}
ds^{2} = -e^{2\Phi(r)}dt^{2}+ \frac{1}{1-\frac{b(r)}{r}}dr^{2}+r^{2}d\Omega^{2}\,,\label{line}
\end{eqnarray}
with $d\Omega^{2}\equiv d\theta^{2}+\sin^{2}\theta d\phi^{2}$. Notice that the above metric was recently extended to a general Morris-Thorne wormhole having an electric charge by considering a spacetime of embedding class I \cite{Kuhfittig:2011xh}. In the wormhole geometry, the redshift function $\Phi(r)$ has to be finite in order to avoid the formation of an event horizon. Additionally, the shape
function $b(r)$ determines the wormhole geometry. In order to have a consistent WH construction, the shape function should satisfy the following properties: ($i$) $b(r)/r<1$ for $r>r_{0}$,\,($ii$) $b(r)=r_{0}$ at $r=r_{0}$,\,($iii$) $b(r)/r\rightarrow 0$ as $r\rightarrow \infty$,\,($iv$) $b'(r)r-b(r)<0$ and ($v$) $b'(r)<1$ at $r=r_{0}$. The Einstein–Maxwell equations for static charged Casimir wormholes can be parametrized in terms of density $\rho(r)$, radial pressure $p_r$,
tangential pressure $p_t$ and electric charge $E(r)=Q/r^{2}$. Here the electromagnetic stress–energy tensor is given by $T_{\mu\nu}^{EM}=\tfrac{1}{4\pi}(F_{\mu\lambda}F^{\lambda}_{\,\,\,\,\nu}-\tfrac{1}{4}g_{\mu\nu}F_{\lambda\sigma}F^{\lambda\sigma})$ and $F_{\mu\nu}$ represents the electromagnetic strength tensor. With the help of the line element given in Eq.(\ref{line}), we obtain the following set of equations resulting from the energy-momentum components to yield \cite{Maurya:2020l}
\begin{eqnarray}
\frac{1}{r^{2}}-e^{-\lambda}\left[\frac{1}{r^{2}}-\frac{\lambda'}{r}\right] &=& 8\pi\rho+E^{2}\,,\\ -\frac{1}{r^{2}} +e^{-\lambda}\left[\frac{1}{r^{2}}+\frac{\nu'}{r}\right] &=& 8\pi p_{r}-E^{2}\,,\\\frac{e^{-\lambda}}{4}\left[2\nu''+\nu'^{2}-\lambda'\nu' +\frac{2(\nu'-\lambda')}{r}\right] &=& 8\pi p_{t}+E^{2}\,.
\end{eqnarray}
where $\nu(r)\equiv 2\Phi(r)$ and $\exp(\lambda)\equiv(1-b(r)/r)^{-1}$. In terms of $\Phi,\,Q$ and $b(r)$, they read
\begin{eqnarray}
\frac{b'(r)}{r^{2}} &=& 8\pi \rho^{\text{eff}}\,,\label{br}\\ 2\Big(1-\frac{b(r)}{r}\Big)\frac{\Phi'}{r} -\frac{b(r)}{r^{3}} &=& 8\pi p_r^{\text{eff}}\,,\label{2eq}\\\Big(1-\frac{b(r)}{r}\Big)\bigg[\Phi''+(\Phi')^{2}-\frac{b'(r)r-b(r)}{2r(r-b(r))}\Phi'-\frac{b'(r)r-b(r)}{2r^{2}(r-b(r))} +\frac{\Phi'}{r}\bigg] &=& 8 \pi p_t^{\text{eff}}\,.\label{3eq}
\end{eqnarray}
Here $p_{t}=p_{\theta}=p_{\phi}$ is assumed. For convenience, we have defined effective parameters $\rho^{\rm eff} \equiv \rho + \tfrac{Q^{2}}{8\pi r^{4}}\,, p_r^{\rm eff} \equiv p_{r}-\tfrac{Q^{2}}{8 \pi r^{4}}$ and $p_t^{\rm eff} \equiv p_{t} + \tfrac{Q^{2}}{8\pi r^{4}}$.

\subsection{Charged Casimir wormholes without GUP}
\label{noGUP}
The Casimir effect was predicted theoretically by Hendrik B. G. Casimir \cite{Casimir:1948dh} in 1948. The Casimir force is a time-honored example of the mechanical effect of vacuum fluctuations. It is an interaction
of a pair of neutral, parallel conducting planes caused by the vacuum fluctuations of the electromagnetic
field, and however only
in recent years reliable experimental investigations have confirmed such a phenomenon \cite{Lamoreaux:1996wh}. The Casimir effect is a macroscopic quantum
effect which causes the plates to attract each other by negative energy. It was found that the energy per unit surface
is given 
\begin{eqnarray}
{\cal E} = -\frac{\pi^{2}\hbar c}{720 a^{3}}\,,
\end{eqnarray}
where $a$ is a distance between plates along the $z$-axis, the direction perpendicular to the plate. The full expression of the energy-momentum tensor for the Casimir effect and a suitable form applying in TW can be found in Ref.\cite{Garattini:2019ivd} and we do not repeat it here. From now on we will consider $\hbar=1=c$, unless otherwise stated. Consequently, the finite force per unit area acting between the plates can be determined to obtain ${\cal F}=-d{\cal E}(a)/da=-3{\cal E}(a)/a$ producing also a pressure of the form $p=-3\pi^{2}/720a^{3}$. Defining an equation-of-state (EoS) parameter $\omega=p/\rho$, in the case of Casimir energy there is a natural EoS establishing fundamental relationship by choosing $\omega=3$. Therefore, we obtain the energy density of the form $\rho=-\pi^{2}/720a^{3}$. It has been proven in Refs.\cite{Boyer:1968uf} that there is an existence of the positive energy of the Casimir effect in a conducting spherical shell with the radius $r$ leading to a thermodynamic instability. Consequently, the author of \cite{Garattini:2019ivd} has extended the study began by Morris, Thorne and Yurtsever in Ref.\cite{Morris:1988cz} and subsequently
explored by Visser \cite{Visser1995} on the nagative Casimir energy as a possible source for constructing a TW. For more detail discussions and mathematical derivations on the Casimir energy in TW, we refer to Ref. \cite{Visser1995}.

In this section, we use the Casimir energy density to figure out the shape function $b(r)$. We are also interested in deriving the equation of state allowing to connect pressure with energy density for a given wormhole geometry. In other words, we fix the geometry parameters using different redshift functions $\Phi$ and then
examine what the EoS parameter in the corresponding case
is. Having assumed a variable separation between the plates, we write the energy density as $\rho=-\pi^{2}/720r^{4}$.

\subsubsection{$\Phi(r)={\rm const.}$}
\begin{figure}[!h]	
	\includegraphics[width=7.5cm]{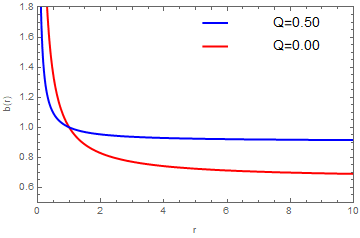}
	\includegraphics[width=7.5cm]{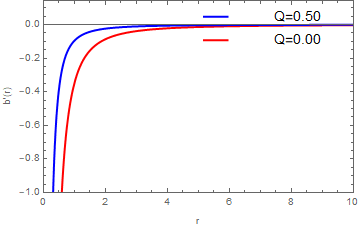}
	\includegraphics[width=7.5cm]{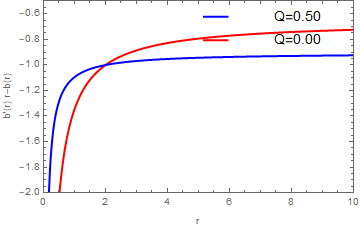}
	\centering
	\caption{We show behaviors of $b(r)$ (upper left),\,$b'(r)$ (upper right) and $b'(r)r-b(r)$ (lower one) without GUP corrections of charged Casimir wormhole against $r$. We have used $r_{0}=1.0$, $Q=0.0$ (red), $Q=0.25$ (blue) and $Q=0.50$ (green).}
	\label{br1}
\end{figure}

We begin with the simplest case in which a model is described by $\Phi={\rm const.}$ \cite{Morris:1988cz}, a spacetime with no tidal forces, namely $\Phi'(r)=0$. In other words, this is asymptotically flat wormhole spacetime. From (\ref{br}), we can simply solve to obtain
\begin{eqnarray}
b(r)= r_{0}+\frac{\pi^3-90 Q^2}{90} \left(\frac{1}{r}-\frac{1}{r_0}\right),\label{brq}
\end{eqnarray}
where the throat condition $b(r_{0})=r_{0}$ has been imposed. Using a condition ($i$), upper-bound values of $Q$ are constrained to be
\begin{eqnarray}
r>r_{0}\land |Q|<\frac{\sqrt{90 r r_0+\pi ^3}}{3 \sqrt{10}}\,,\label{Q}
\end{eqnarray}
Notice from Eq.(\ref{Q}) that values of $Q$ increase when $r$ increases. Important behaviors of $b(r)$ are displayed in Fig.(\ref{br1}). The asymptotically flat metric can be seen also from Fig.(\ref{br1}) for $b'(r)$.
The scaling of coordinate so that $\exp(2\Phi)dt^{2}\rightarrow dt^{2}$ is introduced since $\exp(2\Phi)={\rm const.}$. Substituting $b(r)$ into Eq.(\ref{line}), the wormhole metric becomes
\begin{eqnarray}
ds^{2} = -dt^{2}+ \frac{1}{1-\frac{r_{0}}{r}-\frac{\pi^3-90 Q^2}{90r} \left(\frac{1}{r}-\frac{1}{r_0}\right)}dr^{2}+r^{2}d\Omega^{2}\,,\label{line1}
\end{eqnarray}
Clearly, we can simply show that $\lim_{x \to \infty} b(r)/r\rightarrow 0$. Defining the EoS $p^{\rm eff}_{r}(r) = \omega_{r}(r)\rho^{\rm eff}(r)$ \cite{Azreg-Ainou:2014dwa,Moraes:2017dbs}, and assuming $\Phi'(r)=0$, we obtain 
\begin{eqnarray}
-\frac{b(r)}{r^{3}} = 8\pi p_r^{\text{eff}}=8\pi \omega_{r}(r)\rho^{\rm eff}\,.\label{line2}
\end{eqnarray}
We then solve the above equation for the EoS parameter to obtain
\begin{eqnarray}
\omega_{r}(r) = \frac{r_0 \left(-180 Q^2+90 r r_0+\pi ^3\right)-\left(\pi ^3-90 Q^2\right) r}{\pi ^3 r_0}\,.
\end{eqnarray}
\begin{figure}[!h]	
	\includegraphics[width=9cm]{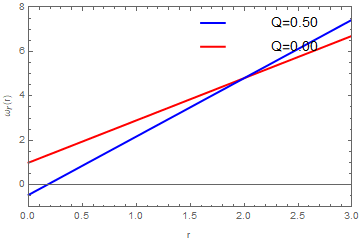}
	\centering
	\caption{We show behaviors of $\omega_{r}(r)$ of charged Casimir wormhole without GUP corrections against $r$. We have used $r_{0}=1.0$, $Q=0.0$ (red), and $Q=0.50$ (blue).}
	\label{ome112}
\end{figure}
The behavior of $\omega_{r}(r)$ of charged Casimir wormhole against $r$ is displayed in Fig.(\ref{ome112})

\subsubsection{$\Phi(r)=\frac{r_{0}}{r}$}
In this form of $\Phi(r)$ suggested by Ref.\cite{Rahaman2019}, we do mainly focus on deriving the EoS parameters. Here for given $\Phi$ we consider $p_{r}$ and $p_{t}$.

\subsubsection*{EoS $p^{\rm eff}_{r}(r)=\omega_{r}(r)\rho^{\rm eff}(r)$}
In this subsection, let's first define the EoS parameter written in the form $p^{\rm eff}_{r}(r)=\omega_{r}(r)\rho^{\rm eff}(r)$ \cite{Azreg-Ainou:2014dwa,Moraes:2017dbs}. From the Einstein's field
given in Eq.(\ref{2eq}), then we solve the resulting relation to obtain
\begin{eqnarray}
    \omega _r(r)&=&\frac{1}{\pi ^3 r r_0}\bigg[r_0 \bigg(3 \left(\pi ^3-120 Q^2\right) r-2 r_0 \left(-45 \left(2 Q^2+3 r^2\right)+90 r r_0+\pi ^3\right)\bigg)\nonumber\\&&\quad\quad\quad\quad\quad\quad\quad\quad-\left(\pi ^3-90 Q^2\right) r^2\bigg]\,.
\end{eqnarray}

\subsubsection*{EoS $p^{\rm eff}_{t}(r)=\omega_{t}(r)p^{\rm eff}_{r}(r)$}
In the second type of the EoS parameter, we consider the case in which $p^{\rm eff}_{t}(r)=\omega_{t}(r)p^{\rm eff}_{r}(r)$ \cite{Azreg-Ainou:2014dwa,Moraes:2017dbs}, where $\omega_{t}(r)$ is as an arbitrary function of $r$. In this case, we combine Eq.(\ref{2eq}) with Eq.(\ref{3eq}) and find the following relation:
\begin{eqnarray}
&&r \bigg(\left(r \Phi '(r)+1\right) \left(-r b'(r)+2 r (r-b(r)) \Phi '(r)+b(r)\right)+2 r^2 (r-b(r)) \Phi ''(r)\bigg)\nonumber\\&&-2 \bigg(\omega _t \big(2 r^2 (r-b(r)) \Phi '(r)-r b(r)+Q^2\big)+Q^2\bigg)=0\,.
\end{eqnarray}
Again we are going to use the shape function (\ref{brq}) for the EoS parameter and then we obtain
\begin{eqnarray}
\omega_{t}(r)=\frac{r^2 \left(\alpha _1-\pi ^3 \left(r-2 r_0\right)\right)+r_0 r \left(\pi ^3 \left(3 r-4 r_0\right)-\alpha _3\right)+{\cal B}}{2 r \left(r \left(\pi ^3 \left(r-r_0\right)-\alpha _1\right)-2 \alpha _2 \left(r-r_0\right) r_0\right)}\,,
\end{eqnarray}
where we have defined new parameters
\begin{eqnarray}
{\cal B}&=&2 \alpha _2 \left(r-r_0\right) r_0^2\,,\nonumber\\
\alpha _1&=&90 \left(Q^2 (r-2 r_0)+r r_0^2\right)\,,\nonumber\\
\alpha _2&=&-90 Q^2+90 r r_0+\pi ^3\,,\nonumber\\
\alpha _3&=&90 \left(Q^2 (3 r-4 r_0)+r r_0 (3 r_0-2 r)\right)\,.\nonumber
\end{eqnarray}

\begin{figure}[!h]	
\begin{center}
\includegraphics[width=7.5cm]{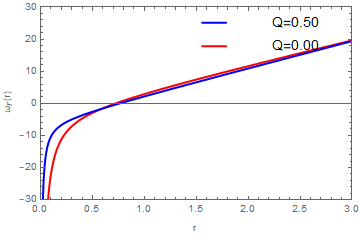}
\includegraphics[width=7.5cm]{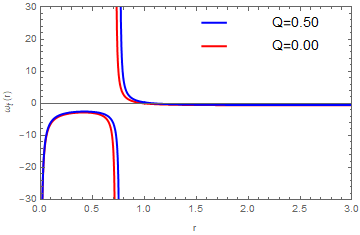}
\caption{We show the EoS parameterr of $\omega_{r}(r)$ (left panel) and $\omega_{t}(r)$ (right panel) of charged Casimir wormhole against $r$ for a non-constant redshift function $\Phi(r)=\tfrac{r_{0}}{r}$. We have used $r_{0}=1.0$, $Q=0.0$ (red), and $Q=0.50$ (blue). \label{omert}}
\end{center}
\end{figure}
We display the behavior of the EoS parameterr $\omega_{t}(r)$ of charged Casimir wormhole charaterized in Fig.(\ref{omert})

\subsubsection{$\Phi(r)=\tfrac{1}{2}\ln\left(1+\frac{\gamma^{2}}{r^{2}}\right)$}
Another interesting example is the following wormhole metric given by
\begin{eqnarray}
ds^{2}=-\bigg(1+\frac{\gamma^{2}}{r^{2}}\bigg)dt^{2}+\frac{1}{1-\frac{b(r)}{r}}dr^{2}+r^{2}d\Omega^{2}\,,
\end{eqnarray}
where $\gamma$ is some positive parameter with $r \geq r_{0}$. This form of $\Phi$ was proposed in Ref.\cite{Rahaman2019}. As we presented in the
preceeding subsection, we can also assume the EoS of the form $p^{\rm eff}_{r}(r) = \omega_{r}(r)\,\rho^{\rm eff}(r)$ \cite{Azreg-Ainou:2014dwa,Moraes:2017dbs}. Similarly, in this case, we find
\begin{eqnarray}
\omega _r(r) &=& \frac{1}{3 \pi ^3 r^2 r_0^3 \left(\gamma ^2+r^2\right)}\Big(270 r^3 r_0^2 \left(r \left(Q^2 \left(r-2 r_0\right)+r r_0^2\right)-\gamma ^2 \left(Q^2+r_0 \left(r_0-2 r\right)\right)\right)\nonumber\\&&\quad\quad\quad\quad\quad\quad\quad\quad-3 \pi^3 r^2 \left(r-r_0\right) r_0^2 (r-\gamma ) (\gamma +r)\Big)\,.
\end{eqnarray}
In the same manner with the preceding scenario, we in this case consider the EoS parameter which is of the form $p^{\rm eff}_{t}(r)=\omega_{t}(r)\,p^{\rm eff}_{r}(r)$ \cite{Azreg-Ainou:2014dwa,Moraes:2017dbs}, where $\omega_{t}(r)$ is as an arbitrary function of $r$. In this case, combining Eq.(\ref{2eq}) with Eq.(\ref{3eq}), we find the following
equation:
\begin{eqnarray}
\omega_t(r) =\frac{r b(r) \left(\gamma ^2 r_0 \left(\gamma ^2+5 r^2+2 \gamma ^2 r_0\right)-\left(\gamma ^2+r^2\right)^2\right)+{\cal H}_1-2 \gamma ^4 r^2 r_0^2}{2 \left(\gamma ^2+r^2\right) \left(\left(\gamma ^2+r^2\right) \left(r b(r)-Q^2\right)+2 \gamma ^2 r r_0 (r-b(r))\right)}\,,
\end{eqnarray}
where we have defined a new parameter
\begin{eqnarray}
{\cal H}_1=\left(\gamma ^2+r^2\right)^2 \left(r^2 b'(r)+2 Q^2\right)-\gamma ^2 r^2 r_0 \left(\left(\gamma ^2+r^2\right) b'(r)+4 r^2\right)\,.
\end{eqnarray}
\begin{figure}[!h]	
\begin{center}		
\includegraphics[width=7.5cm]{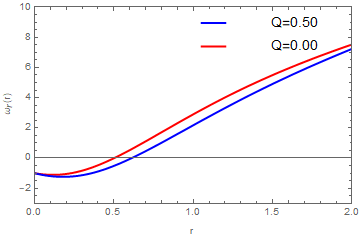}
\includegraphics[width=7.5cm]{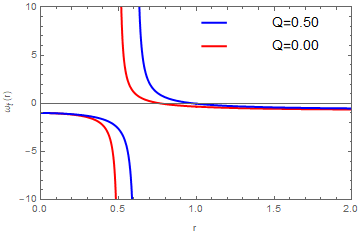}
\caption{We show the EoS parameterr $\omega_{r}(r)$ and $\omega_{t}(r)$ of charged Casimir wormhole against $r$ for a non-constant redshift function $\exp(2\Phi(r))=1+\tfrac{\gamma^{2}}{r^{2}}$. We have used $r_{0}=1.0,\,\gamma=1$, $Q=0.0$ (red), and $Q=0.50$ (blue). \label{g1g}}
\end{center}
\end{figure}
We display the behavior of the EoS parameters $\omega_{r}(r)$  and $\omega_{t}(r)$ of charged Casimir wormhole characterized in Fig.(\ref{g1g})

\subsection{Charged Casimir wormholes with GUP}
\label{GUP}
In the starting point, we will follow the work present in Ref.\cite{Jusufi:2020rpw} for elaborating in more details about the GUP corrected energy density to construct Casimir wormholes. To be more precise, the authors of Ref.\cite{Jusufi:2020rpw} have considered three types of
the GUP relations: (1) the Kempf, Mangano and Mann
(KMM) model, (2) the Detournay, Gabriel and Spindel
(DGS) model, and (3) the so called type II model for GUP
principle. In a compact form, the authors of Ref.\cite{Jusufi:2020rpw} wrote the renormalized energies per unit surface area of the plates for three GUP cases as
\begin{eqnarray}
{\cal E} = -\frac{\pi^{2}\hbar}{720 a^{3}}\bigg[1+C_{i}\bigg(\frac{\hbar\sqrt{\beta}}{a}\bigg)^{2}\bigg]\,,
\end{eqnarray}
where we have defined $C_{i}$ as
\begin{eqnarray}
C_{1}=\pi^{2}\bigg(\frac{28 + 3\sqrt{10}}{14}\bigg),\,C_{2}= 4\pi^{2}\bigg(\frac{3+\pi^{2}}{21}\bigg),\,C_{3}=\frac{2\pi^{2}}{3}\,.
\end{eqnarray}
Therefore, the force per unit surface area can be determined to obtain
\begin{eqnarray}
{\cal F} = -\frac{d{\cal E}}{da} = -\frac{3\pi^{2}\hbar}{720 a^{4}}\bigg[1+C_{i}\bigg(\frac{\hbar\sqrt{\beta}}{a}\bigg)^{2}\bigg]\,.
\end{eqnarray}
Hence the pressure can be simply obtained as
\begin{eqnarray}
p={\cal F} = -\frac{3\pi^{2}\hbar}{720 a^{4}}\bigg[1+C_{i}\bigg(\frac{\hbar\sqrt{\beta}}{a}\bigg)^{2}\bigg]\,.
\end{eqnarray}
By defining an EoS, $p=\omega\rho$, we obtain the GUP corrected energy density in a compact form as
\begin{eqnarray}
\rho = -\frac{\pi^{2}\hbar}{720 a^{4}}\bigg[1+D_{i}\bigg(\frac{\hbar\sqrt{\beta}}{a}\bigg)^{2}\bigg]\,,
\end{eqnarray}
with $D_{i}=5C_{i}/3$. Notice that in the case of Casimir energy there is a natural EoS establishing fundamental relationship by choosing $\omega=3$. Practically, the energy density can be employed to quantify the shape function $b(r)$. Additionally, the EoS with a specific value for $\omega$ can be used to determine the red-shift function.

\subsubsection{$\Phi(r)={\rm const.}$}
\begin{figure}[!h]	
	\includegraphics[width=9cm]{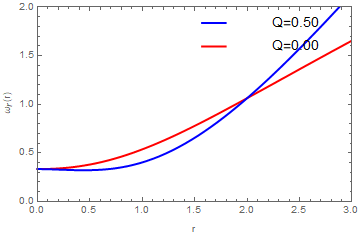}
	\centering
	\caption{We show behaviors of $\omega(r)$ of charged Casimir wormhole with GUP corrections against $r$. We have used $r_{0}=1.0,\,D_{i}=D_{1},\,\beta=0.1$, $Q=0.0$ (red), and $Q=0.50$ (blue).}
	\label{ome1}
\end{figure}

\begin{figure}[!h]	
	\includegraphics[width=7.5cm]{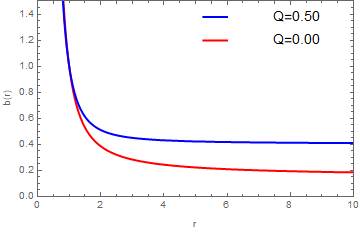}
	\includegraphics[width=7.5cm]{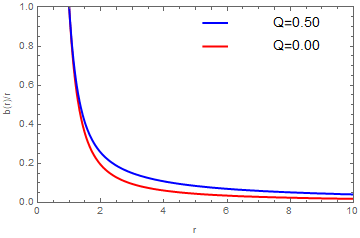}
	\includegraphics[width=7.5cm]{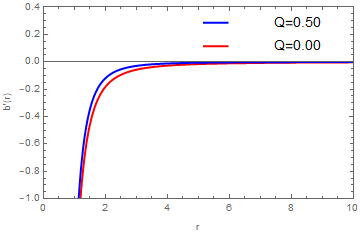}
	\includegraphics[width=7.5cm]{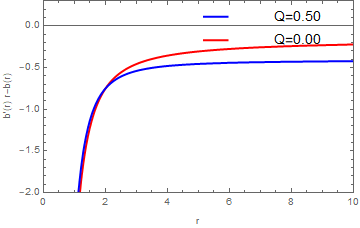}
	\centering
	\caption{We show the behavior of $b(r)$ (upper-left panel),\,$b'(r)$ (lower-left panel),\,$b(r)/r$ (upper-right panel) and $b'(r)r-b(r)$ (lower-right panel) of charged Casimir wormhole with GUP corrections against $r$ using $\Phi={\rm const.}$. We have used $r_{0}=1.0,\,D_{i}=D_{1}$, $Q=0.0$ (red), and $Q=0.50$ (blue).}
	\label{brqb}
\end{figure}

One of the simplest case is a model with $\Phi={\rm const.}$ \cite{Morris:1988cz}, a.k.a., a spacetime with no tidal forces, namely $\Phi'(r)=0$. In other words, this is asymptotically flat wormhole spacetime. From (\ref{br}), we find
\begin{eqnarray}
b(r)= r_{0}+\frac{\pi^3-90 Q^2}{90} \left(\frac{1}{r}-\frac{1}{r_0}\right)+\frac{\pi ^3 \beta D_{i}}{270}\left(\frac{1}{r^{3}}-\frac{1}{r^{3}_0}\right),\label{brq11}
\end{eqnarray}
where the throat condition $b(r_{0})=r_{0}$ has been imposed. Behavior of $b(r)$ is displayed in Fig.(\ref{brqb}). Using a condition ($i$), we find the conditions
\begin{eqnarray}
r>r_{0}\land|Q|<\frac{1}{3 \sqrt{30}}\bigg[\sqrt{\pi^3 \left(\frac{\beta D_{i}}{r^2}+3\right)+\frac{\pi^3 \beta D_{i}}{r r_0}+\frac{\pi^3 \beta D_{i}}{r_0^2}+270 r r_0}\bigg]\,.
\end{eqnarray}
Solving equation for the EoS parameter, we obtain in this case
\begin{eqnarray}
\omega(r)= \frac{-\pi^3 \beta D_{i} r^3+\pi^3 \beta D_{i} r_0^3+270 Q^2 r^3 r_0^2-270 Q^2 r^2 r_0^3+{\cal F}}{3 \pi^3 r_0^3 \left(\beta D_{i}+r^2\right)}\,,
\end{eqnarray}
where we have defined a new parameter
\begin{eqnarray}
{\cal F}=270 r^3 r_0^4-3 \pi ^3 r^3 r_{0}^2+3 \pi^3 r^2 r_0^3\,.
\end{eqnarray}
The behavior of $\omega_{r}(r)$ of charged Casimir wormhole with GUP corrections against $r$ is shown in Fig.\ref{ome1}.

\subsubsection{$\Phi(r)=\frac{r_{0}}{r}$}
\begin{figure}[!h]	
\begin{center}		
\includegraphics[width=7.5cm]{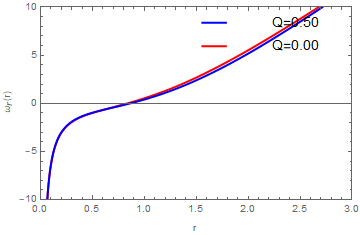}
\includegraphics[width=7.5cm]{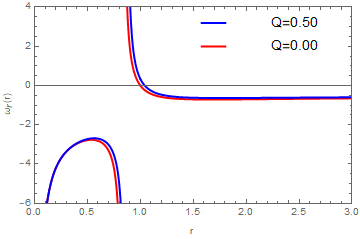}
\caption{We show the EoS parameter $\omega_{r}(r)$ and $\omega_{t}(r)$ of charged Casimir wormhole against $r$ for a non-constant redshift function $\Phi(r)=r_{0}/r$. We have used $r_{0}=1.0,\,\beta=0.1,\,D_{i}=D_{1}$, $Q=0.0$ (red), and $Q=0.50$ (blue). \label{g1}}
\end{center}
\end{figure}

In this subsection, we first define the EoS $p^{\rm eff}_{r}(r)=\omega_{r}(r)\rho^{\rm eff}(r)$ \cite{Azreg-Ainou:2014dwa,Moraes:2017dbs}. Then from the Einstein's field
equations (\ref{2eq}), we come up with
\begin{eqnarray}
\frac{rb(r)-\frac{1}{90} \pi ^3 \omega _r \left(\frac{\beta  D_{i}}{r^2}+1\right)-Q^2}{2 r^2 \left(-\frac{1}{270} \pi ^3 \beta D_{i} \left(\frac{1}{r^3}-\frac{1}{r_0^3}\right)-\frac{1}{90} \left(\pi^3-90 Q^2\right) \left(\frac{1}{r}-\frac{1}{r_0}\right)+r-r_0\right)}+\frac{r_0}{r^2}=0\,.
\end{eqnarray}
We then solve the above equation on yield
\begin{eqnarray}
\omega _r = \frac{r_0 \left(2 \pi^3 \beta D_{i} r^3+r_0 \left(G_1 r_0-3 \left(\pi^3-90 Q^2\right) r^4\right)\right)-\pi^3 \beta D_{i} r^4}{3 \pi^3 r_0^3 \left(\beta  D_{i} r+r^3\right)}\,,
\end{eqnarray}
where we have defined a new parameter
\begin{eqnarray}
G_1 &=&-2 r_0 \left(\pi^3 \beta D_{i}+3 r^2 \left(\pi^3-45 \left(2 Q^2+3 r^2\right)\right)+270 r_0 r^3\right)\nonumber\\&&+\pi^3 \beta  D_{i} r+9 \left(\pi^3-120 Q^2\right) r^3\,.\nonumber
\end{eqnarray}
We can follow the basic formalism given in the preceding section and consider the scenario in which the EoS parameter is written in the form $p^{\rm eff}_{t}(r)=\omega_{t}(r)\,p^{\rm eff}_{r}(r)$ \cite{Azreg-Ainou:2014dwa,Moraes:2017dbs}, where $\omega_{t}(r)$ is as an arbitrary function of $r$. In this case, we obtain
\begin{eqnarray}
\omega _t = \frac{r \left(r \left(r-r_0\right) b'(r)+2 Q^2-2 r_0 \left(r+r_0\right)\right)+\left(-r^2+3 r_0 r+2 r_0^2\right) b(r)}{2 r \left(r b(r)+2 r_0 (r-b(r))-Q^2\right)}\,.
\end{eqnarray}
We display the behavior of the EoS parameter $\omega_{r}(r)$ and $\omega_{t}(r)$ of charged Casimir wormhole charaterized in Fig.(\ref{g1}).

\subsubsection{$\Phi(r)=\frac{1}{2}\ln\left(1+\tfrac{\gamma^{2}}{r^{2}}\right)$}
\begin{figure}[!h]	
\begin{center}		
\includegraphics[width=7.5cm]{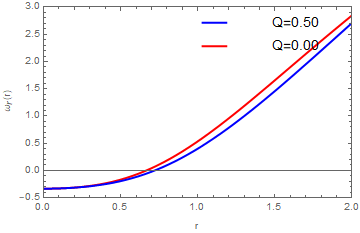}
\includegraphics[width=7.5cm]{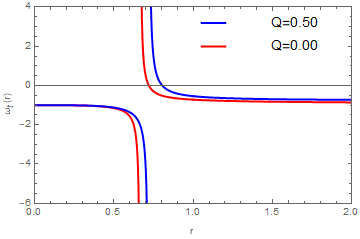}
\caption{We show the EoS parameterr $\omega_{r}(r)$ and $\omega_{t}(r)$ of charged Casimir wormhole against $r$ for a non-constant redshift function $\Phi(r)=\tfrac{1}{2}\ln{(1+\tfrac{\gamma^{2}}{r^{2}})}$. We have used $r_{0}=1.0,\,\beta=0.1,\,D_{i}=D_{1},\,\gamma=1$, $Q=0.0$ (red), and $Q=0.50$ (blue). \label{g13}}
\end{center}
\end{figure}

In the last example of the redshift function, we consider the EoS parameter which is given by $p^{\rm eff}_{r}(r)=\omega_{r}(r)\rho^{\rm eff}(r)$ \cite{Azreg-Ainou:2014dwa,Moraes:2017dbs}. Considering the Einstein's field equations (\ref{2eq}), we obtain
\begin{eqnarray}
\omega _r = \frac{\pi ^3 \beta D_{i} r^3 \left(\gamma ^2-r^2\right)+r_0^2 \left(F_1 r_0-3 \left(\pi^3-90 Q^2\right) r^3 (r-\gamma ) (\gamma +r)\right)}{3 \pi^3 r_0^3 \left(\gamma ^2+r^2\right) \left(\beta D_{i}+r^2\right)}\,,
\end{eqnarray}
where we have defined a new parameter
\begin{eqnarray}
F_1 =\pi ^3 (r-\gamma ) (\gamma +r) \left(\beta D_{i}+3 r^2\right)+540 r^4 \left(\gamma ^2-Q^2\right)+270 r_0 r^3 (r-\gamma ) (\gamma +r)\,.\nonumber
\end{eqnarray}
Let us now define the EoS of the form $p^{\rm eff}_{t}(r)=\omega_{t}(r)\,p^{\rm eff}_{r}(r)$ \cite{Azreg-Ainou:2014dwa,Moraes:2017dbs}, where $\omega_{t}(r)$ is as an arbitrary function of $r$. In this case, we find the following
relation:
\begin{eqnarray}
\omega _t = -\frac{r^4 \left(\gamma ^2+r^2\right) b'(r)+b(r) \left(-r^5+3 \gamma ^2 r^3+2 \gamma ^4 r\right)+{\cal O}_{1}}{2 \left(\gamma ^2+r^2\right) \left(r b(r) \left(\gamma ^2-r^2\right)+Q^2 \left(\gamma ^2+r^2\right)-2 \gamma ^2 r^2\right)}\,,
\end{eqnarray}
where we have defined ${\cal O}_{1}=2 Q^2 \left(\gamma ^2+r^2\right)^2-2 \gamma ^2 r^2 \left(\gamma ^2+2 r^2\right)$. We display the behavior of the EoS parameter $\omega_{r}(r)$ and $\omega_{t}(r)$ of charged Casimir wormhole charaterized in Fig.(\ref{g13}).

\section{Embedding diagram}
\label{embed}
\begin{figure}[!h]	
\begin{center}		
\includegraphics[width=7.5cm]{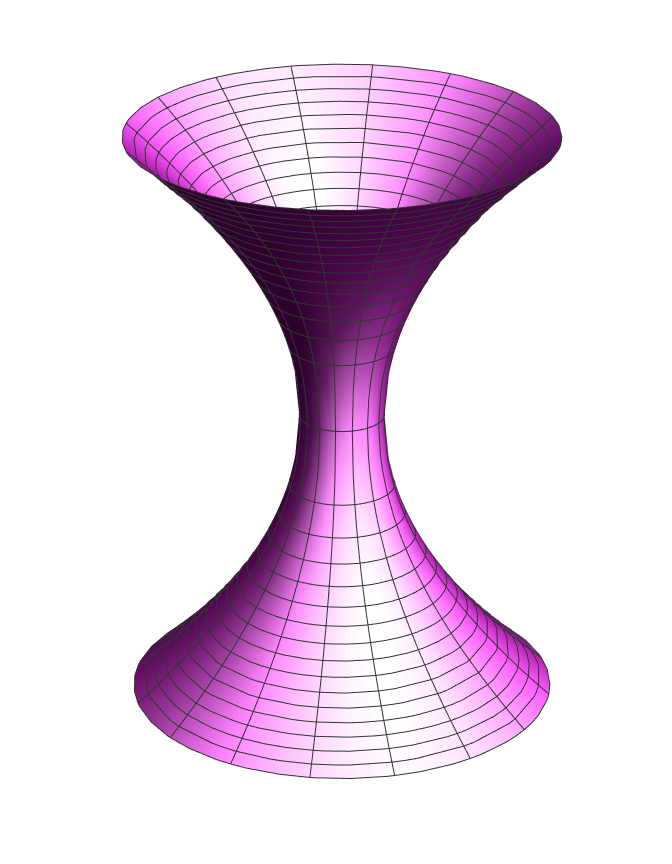}
\includegraphics[width=7.5cm]{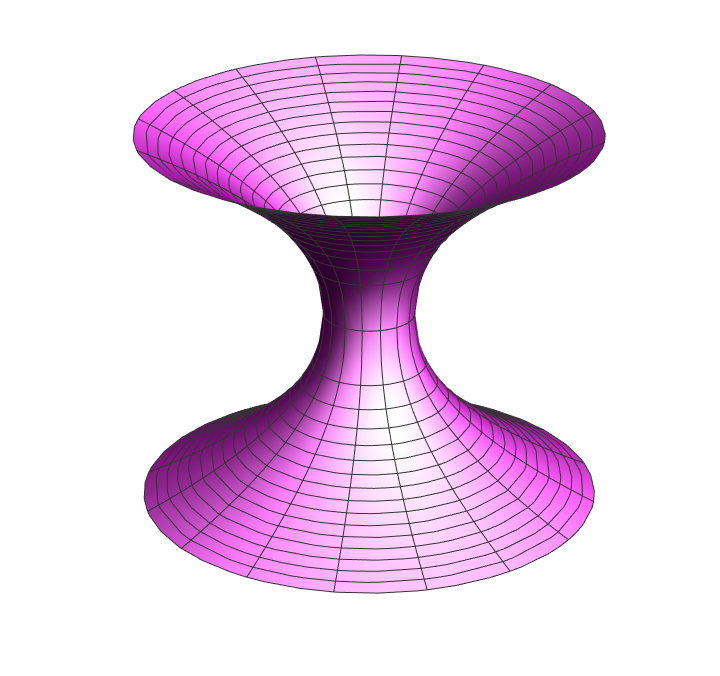}
\caption{The charged Casimir wormhole without GUP corrections (left panel) and with GUP corrections (right panel) embedded in a three-dimensional Euclidean space. Left panel: We have used $r_{0}=1,\,Q=0.25$ for $b(r)$ given in Eq.(\ref{brq}). Right
panel: We have used $r_{0}=1,\,Q=0.25,\,\beta=0.1,\,D_{i}=D_{1}$ for $b(r)$ given in Eq.(\ref{brq11}).
\label{gmod3}}
\end{center}
\end{figure}

In this subsection, we analyse the embedding diagrams to
represent the charged Casimir wormhole without and with GUP corrections by considering an equatorial slice $\theta=\pi/2$ at some fix moment
in time $t={\rm constant}$. To do so, we consider the metric which can be written as
\begin{eqnarray}
ds^{2}=\frac{dr^{2}}{1-\frac{b(r)}{r}}+r^{2}d\phi^{2}\,.
\end{eqnarray}
We embed the metric (\ref{line}) into three-dimensional Euclidean space to visualize this slice and the spacetime
can be written in cylindrical coordinates as
\begin{eqnarray}
ds^{2}=dz^{2}+dr^{2}+r^{2}d\phi^{2}\,.
\end{eqnarray}
From the last two equations we find that
\begin{eqnarray}
\frac{dz}{dr}=\pm \sqrt{\frac{r}{r-b(r)}-1}\,,
\end{eqnarray}
where $b(r)$ is given by Eq.(\ref{brq}) and Eq.(\ref{brq11}). Invoking numerical techniques allows us to illustrate the wormhole shape given in Fig.\ref{gmod3}.

\section{Energy conditions}
\label{ec} 
We can continue our discussion on the issue of energy conditions and make some regional plots to check the validity of all energy conditions. In this work, we consider the three types of energy conditions to examine charged wormholes. The first one is null energy condition (NEC) which determines the non-negative value of energy
momentum tensor contracting with null vector $k_{\mu}$ where $k^{\mu}k_{\mu}=0$. That is to say $T_{\mu\nu}k^{\mu}k^{\nu}\geq 0$. We have
\begin{eqnarray}
    \rho^{\text{eff}} + p_r^{\text{eff}} \geq 0\,. \nonumber
\end{eqnarray}
Note that NEC can be interpreted as the energy of particles traveling
along a null geodesic, such as photon and massless particles, which must be non-negative. The energy density or pressure can be negative as long as their summation is still equal or greater than zero. In particular we recall that the weak energy condition (WEC) is defined by $T_{\mu\nu}t^{\mu}t^{\nu}\geq 0$. It determines the non-negative value of energy momentum tensor contracting with timelike vector $t_{\mu}$ where $t^{\mu}t_{\mu}<0$, i.e.,
\begin{eqnarray}
    \rho^{\text{eff}} + p_{r}^{\text{eff}} \geq 0\,. \nonumber
\end{eqnarray}
The strong energy condition (SEC) decomposed into the following conditions:
\begin{eqnarray}
\rho^{\text{eff}} + 2p_{t}^{\text{eff}} \geq 0\quad{\rm and}\quad \rho^{\text{eff}} + p_{r}^{\text{eff}}+ 2p_{t}^{\text{eff}} \geq 0\,.
\end{eqnarray}
For the sake of completeness, we also need to test the dominant energy condition (DEC) given by 
\begin{eqnarray}
\rho^{\text{eff}} - |p_{r}^{\text{eff}}| \geq 0 \quad{\rm and}\quad \rho^{\text{eff}}  - |p_{t}^{\text{eff}}| \geq 0\,.
\end{eqnarray}
Notice that SEC covers NEC and avoids excessively large negative pressure. However, the traversable wormholes in some particular models \cite{Morris:1988cz} need the exotic matter which violates the energy conditions.

\subsection{Casimir wormholes without GUP}
\label{GUPeo}

\subsubsection{$\Phi(r)=\frac{r_{0}}{r}$}

Given the redshift function $\Phi=r_{0}/r$ and the shape function given Eq.(\ref{brq}), we can compute the energy-momentum components. We find for the radial component 
\begin{eqnarray}
p^{\rm eff}_{r}(r)&=&\frac{1}{720 \pi  r^5 r_0}\bigg[\left(\pi ^3-90 Q^2\right) r^2+r_0 \big(2 r_0 \big(-45 \left(2 Q^2+3 r^2\right)\nonumber\\
&& +90 r r_0+\pi ^3\big)-3 \left(\pi ^3-120 Q^2\right) r\big)\bigg]\,,
\end{eqnarray}
and for the tangential component
\begin{eqnarray}
p^{\rm eff}_{t}(r)=-\frac{C_3 (r-1) \left(C_2 r_0+\left(\pi ^3-90 Q^2\right) r^2\right)}{C_1}\,,
\end{eqnarray}
where
\begin{eqnarray}
C_1&=&1440 \pi  r^6 \left(\pi ^3 (r-2) (r-1)+90 r (2-3 r)\right) r_0\,,\nonumber\\
C_2&=&2 r_0 \left(-45 \left(2 Q^2+3 r^2\right)+90 r r_0+\pi ^3\right)-3 \left(\pi ^3-120 Q^2\right) r\,,\nonumber\\
C_3&=&\pi ^3 ((r-4) r-2)-90 r (3 r+2)\,.\nonumber
\end{eqnarray}
\begin{figure}[!h]	
\begin{center}		
\includegraphics[width=7.5cm]{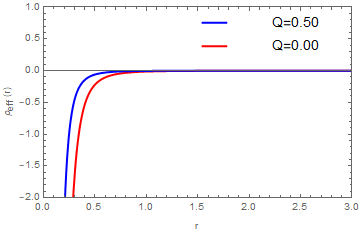}
\includegraphics[width=7.5cm]{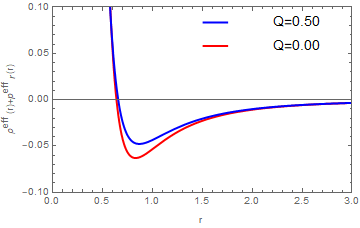}
\includegraphics[width=7.5cm]{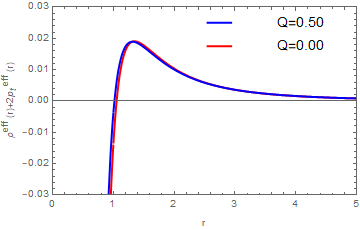}
\includegraphics[width=7.5cm]{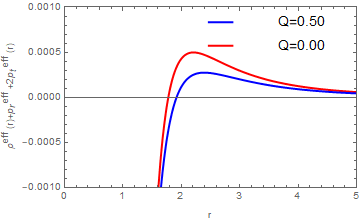}
\includegraphics[width=7.5cm]{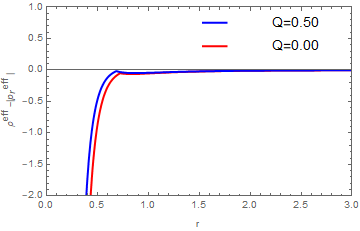}
\includegraphics[width=7.5cm]{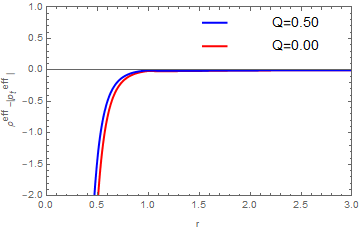}
\caption{We show the variation of $\rho^{\rm eff}(r),\,\rho^{\rm eff}(r)+p^{\rm eff}_{r}(r),\,\rho^{\rm eff}(r)+2p^{\rm eff}_{t}(r),\,\rho^{\rm eff}(r)+p^{\rm eff}_{r}(r)+2p^{\rm eff}_{t}(r),\,\rho^{\rm eff}(r)-|p^{\rm eff}_{r}(r)|$ and $\rho^{\rm eff}(r)-|p^{\rm eff}_{t}(r)|$ as a function of $r$ using $\Phi=r_{0}/r$ of charged Casimir wormhole. We have used $r_{0}=1.0,\,Q=0.0$ (red), and $Q=0.50$ (blue). \label{e10}}
\end{center}
\end{figure}

\subsubsection{$\Phi(r)=\frac12\ln\left(1+\tfrac{\gamma^{2}}{r^{2}}\right)$}
In this case, we compute the energy-momentum components to obtain the radial component as 
\begin{eqnarray}
p^{\rm eff}_{r}(r)=\frac{D_0-r_0 \left(D_1+\pi ^3 (r-\gamma ) (\gamma +r)\right)}{720 \pi  r^4 r_0 \left(\gamma ^2+r^2\right)}\,,
\end{eqnarray}
where
\begin{eqnarray}
D_0&=&\left(\pi ^3-90 Q^2\right) r (r-\gamma ) (\gamma +r)\,,\nonumber\\
D_1&=&180 r^2 \left(\gamma ^2-Q^2\right)+90 r_0 r (r-\gamma ) (\gamma +r)\,,\nonumber
\end{eqnarray}
and the tangential component
\begin{eqnarray}
p^{\rm eff}_{t}(r)=\frac{E_2 \left(E_3 r b(r)-2 \gamma ^4 r^2 r_0^2+E_4 r_0+e_5\right)}{16 \pi E_1 r^4 \left(\gamma ^2+r^2\right)^2}\,,
\end{eqnarray}
where
\begin{eqnarray}
E_0&=&180 r^2 (Q-\gamma ) (\gamma +Q)+90 r_0 r \left(\gamma ^2-r^2\right)+\pi ^3 \left(\gamma ^2-r^2\right)\,,\nonumber\\
E_1&=&\left(90 \left(Q^2 \left(r-2 r_0\right)+r r_0^2\right)+\pi ^3 \left(r_0-r\right)\right) \left(\gamma ^2+r^2\right)+2 \gamma ^2 \left(r-r_0\right) r_0 \left(-90 Q^2+90 r r_0+\pi ^3\right)\,,\nonumber\\
E_2&=&\left(\pi^3-90 Q^2\right) r (r-\gamma ) (\gamma +r)+E_0 r_0\,,\nonumber\\
E_3&=&\gamma ^2 r_0 \left(\gamma^2+5 r^2+2 \gamma^2 r_0\right)-\left(\gamma ^2+r^2\right)^2\,,\nonumber\\
E_4&=&\frac{1}{90} \gamma^2 \left(\pi^3 \left(\gamma ^2+r^2\right)-90 \left(Q^2 \left(\gamma ^2+r^2\right)+4 r^4\right)\right)\,,\nonumber\\
E_5&=&-\frac{1}{90}\left(\pi^3-270 Q^2\right) \left(\gamma ^2+r^2\right)^2\,.\nonumber
\end{eqnarray}

\begin{figure}[!h]	
\begin{center}		
\includegraphics[width=7.5cm]{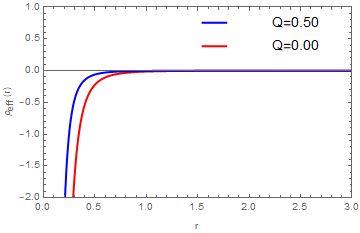}
\includegraphics[width=7.5cm]{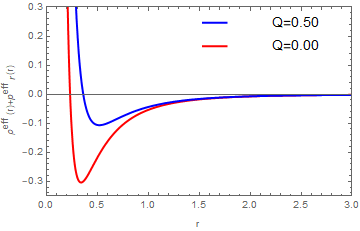}
\includegraphics[width=7.5cm]{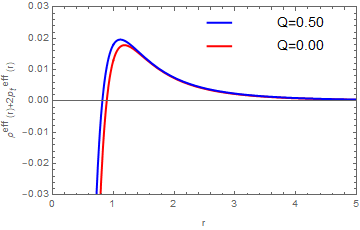}
\includegraphics[width=7.5cm]{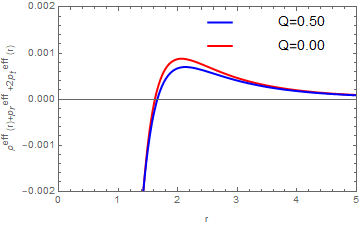}
\includegraphics[width=7.5cm]{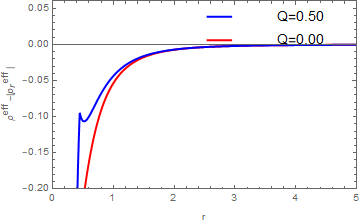}
\includegraphics[width=7.5cm]{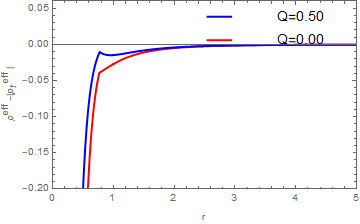}
\caption{We show the variation of $\rho^{\rm eff}(r),\,\rho^{\rm eff}(r)+p^{\rm eff}_{r}(r),\,\rho^{\rm eff}(r)+2p^{\rm eff}_{t}(r),\,\rho^{\rm eff}(r)+p^{\rm eff}_{r}(r)+2p^{\rm eff}_{t}(r),\,\rho^{\rm eff}(r)-|p^{\rm eff}_{r}(r)|$ and $\rho^{\rm eff}(r)-|p^{\rm eff}_{t}(r)|$ as a function of $r$ using $\exp(2\Phi(r))=1+\tfrac{\gamma^{2}}{r^{2}}$ of charged Casimir wormhole. We have used $r_{0}=1.0\,,\gamma=1$, $Q=0.0$ (red), and $Q=0.50$ (blue). \label{e1phi1mod}}
\end{center}
\end{figure}

We see from Fig.(\ref{e10}) for $\Phi=r_{0}/r$, and similarly Fig.(\ref{e1phi1mod}) for $\Phi(r)=\tfrac{1}{2}\ln\left(1+\tfrac{\gamma^{2}}{r^{2}}\right)$, NEC, WEC, and SEC, are violated at the wormhole throat $r=r_{0}$. In other words, we discover that
in all plots at the wormhole throat $r=r_{0}$, we have for the energy condition
$(\rho^{\rm eff}+p^{\rm eff}_{r})|_{r_{0}=1}<0$, along with the condition $(\rho^{\rm eff}+p^{\rm eff}_{r}+2p^{\rm eff}_{t})|_{r_{0}=1}<0$, by arbitrary small values. Moreover, we find that $\rho^{\rm eff}(r)-|p^{\rm eff}_{r}(r)|<0$ and $\rho^{\rm eff}(r)-|p^{\rm eff}_{t}(r)|<0$ at the WH throat.  However, as mentioned in Ref.\cite{Jusufi:2020rpw} from the quantum field theory's point of view, quantum fluctuations are thought to violate most energy conditions without any restrictions. Therefore, wormholes may be stabilized by this quantum fluctuations.

\subsection{Charged Casimir wormholes with GUP}
\label{GUPe}
\subsubsection{$\Phi(r)=\frac{r_{0}}{r}$}
In this case, we compute the energy-momentum components to obtain the radial component as
\begin{eqnarray}
p^{\rm eff}_r(r) &=& \frac{r_0 \left(2 \pi ^3 \beta D_{i} r^3+r_0 \left(G_1 r_0-3 \left(\pi^3-90 Q^2\right) r^4\right)\right)-\pi ^3 \beta D_{i} r^4}{3 \pi^3 r_0^3 \left(\beta D_{i} r+r^3\right)}\nonumber\\&&\times\bigg[\frac{Q^2}{8 \pi  r^4}-\frac{\pi^2 \left(\frac{\beta D_{i}}{r^2}+1\right)}{720 r^4}\bigg]\,,\label{mod2pr}
\end{eqnarray}
and the tangential component
\begin{eqnarray}
p^{\rm eff}_t(r) &=& \frac{r \left(r \left(r-r_0\right) b'(r)+2 Q^2-2 r_0 \left(r+r_0\right)\right)+\left(-r^2+3 r_0 r+2 r_0^2\right) b(r)}{2 r \left(r b(r)+2 r_0 (r-b(r))-Q^2\right)}p_{r}(r)\,,
\end{eqnarray}
where $p_{r}(r)$ has been already given in Eq.(\ref{mod2pr}) and $G_{1}$ is defined as
\begin{eqnarray}
G_{1}&=&-2 r_0 \left(\pi ^3 \beta  D_{i}+3 r^2 \left(\pi ^3-45 \left(2 Q^2+3 r^2\right)\right)+270 r_0 r^3\right)\nonumber\\&&+\pi ^3 \beta  D_{i} r+9 \left(\pi ^3-120 Q^2\right)r^3\,.
\end{eqnarray}

\begin{figure}[!h]	
\begin{center}		
\includegraphics[width=7.5cm]{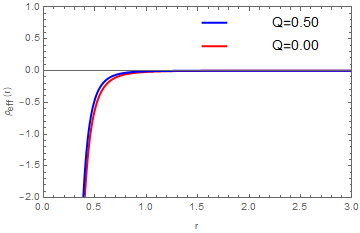}
\includegraphics[width=7.5cm]{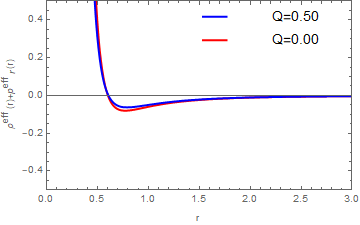}
\includegraphics[width=7.5cm]{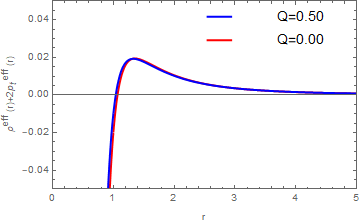}
\includegraphics[width=7.5cm]{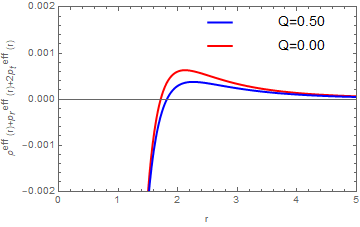}
\includegraphics[width=7.5cm]{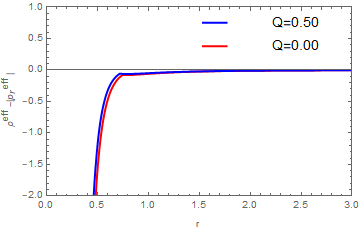}
\includegraphics[width=7.5cm]{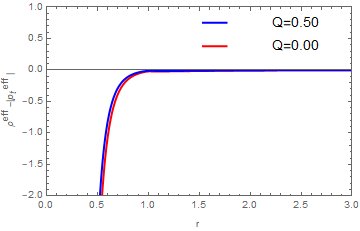}
\caption{We show the variation of $\rho(r)+p_{r}(r),\,\rho(r)+2p_{t}(r)$ and $\rho(r)+p_{r}(r)+2p_{t}(r)$ as a function of $r$ using $\Phi(r)=r_{0}/r$ of charged Casimir wormhole with GUP corrections. We have used $r_{0}=1.0\,,\beta=0.01,\,D_{i}=D_{1}$, $Q=0.0$ (red), and $Q=0.50$ (blue). \label{e1phi1mo}}
\end{center}
\end{figure}

\subsubsection{$\Phi(r)=\frac12\ln\left(1+\tfrac{\gamma^{2}}{r^{2}}\right)$}
Given the redshift function $\exp(2\Phi(r))=1+\tfrac{\gamma^{2}}{r^{2}}$ and the shape function given Eq.(\ref{brq11}), we can compute the energy-momentum components. We find for the radial component 
\begin{eqnarray}
p^{\rm eff}_r(r) &=& \frac{\pi^3 \beta D_{i} r^3 \left(\gamma ^2-r^2\right)+r_0^2 \left(F_1 r_0-3 \left(\pi^3-90 Q^2\right) r^3 (r-\gamma ) (\gamma +r)\right)}{3 \pi^3 r_0^3 \left(\gamma ^2+r^2\right) \left(\beta D_{i}+r^2\right)}\nonumber\\&&\times\bigg(-\frac{\pi ^2 \left(\frac{\beta D_{i}}{r^2}+1\right)}{720 r^4}+\frac{Q^{2}}{8\pi r^{4}}\bigg)\,,
\end{eqnarray}
and for the tangential pressure
\begin{eqnarray}
p^{\rm eff}_t(t) &=& -\frac{r^4 \left(\gamma ^2+r^2\right) b'(r)+b(r) \left(-r^5+3 \gamma ^2 r^3+2 \gamma ^4 r\right)+2 Q^2 \left(\gamma ^2+r^2\right)^2-2 \gamma ^2 r^2 \left(\gamma ^2+2 r^2\right)}{2 \left(\gamma ^2+r^2\right) \left(r b(r) \left(\gamma ^2-r^2\right)+Q^2 \left(\gamma ^2+r^2\right)-2 \gamma ^2 r^2\right)}\nonumber\\&&\times\frac{\pi ^3 \beta D_{i} r^3 \left(\gamma ^2-r^2\right)+r_0^2 \left(F_1 r_0-3 \left(\pi ^3-90 Q^2\right) r^3 (r-\gamma ) (\gamma +r)\right)}{3 \pi ^3 r_0^3 \left(\gamma ^2+r^2\right) \left(\beta D_{i}+r^2\right)}\nonumber\\&&\times\bigg(-\frac{\pi ^2}{720 r^4}\left(\frac{\beta D_{i}}{r^2}+1\right)+\frac{Q^{2}}{8\pi r^{4}}\bigg)\,.
\end{eqnarray}
\begin{figure}[!h]	
\begin{center}		
\includegraphics[width=7.5cm]{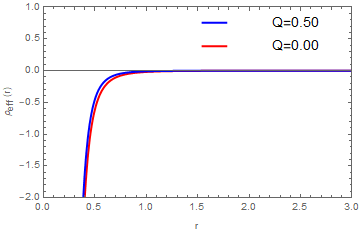}
\includegraphics[width=7.5cm]{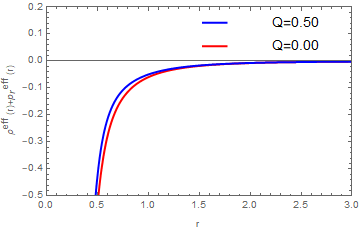}
\includegraphics[width=7.5cm]{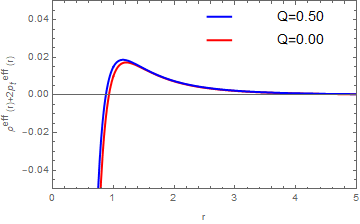}
\includegraphics[width=7.5cm]{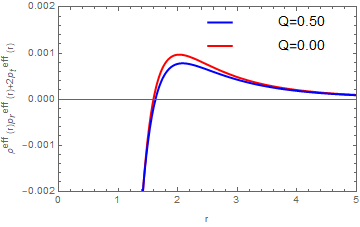}
\includegraphics[width=7.5cm]{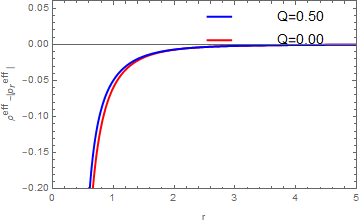}
\includegraphics[width=7.5cm]{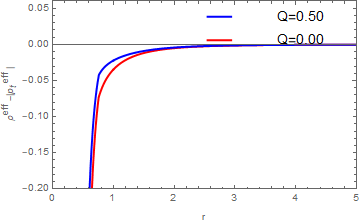}
\caption{We show the variation of $\rho^{\rm eff}(r)+p^{\rm eff}_{r}(r),\,\rho^{\rm eff}(r)+2p^{\rm eff}_{t}(r),\,\rho^{\rm eff}(r)+p^{\rm eff}_{r}(r)+2p^{\rm eff}_{t}(r),\,\rho^{\rm eff}(r)-|p^{\rm eff}_r(r)|$ and $\rho^{\rm eff}(r)-|p^{\rm eff}_t(r)|$ as a function of $r$ using $\exp(2\Phi(r))=1+\gamma^{2}/r^{2}$ of charged Casimir wormhole with GUP corrections. We have used $r_{0}=1.0\,,\beta=0.01,\,D_{i}=D_{1}$, $Q=0.0$ (red), and $Q=0.50$ (blue). \label{e1phi1mod2}}
\end{center}
\end{figure}
In the case of adding the GUP corrections, we also see from Fig.(\ref{e1phi1mo}) for $\Phi=r_{0}/r$, and similarly Fig.(\ref{e1phi1mod2}) for $\Phi(r)=\tfrac{1}{2}\ln\left(1+\tfrac{\gamma^{2}}{r^{2}}\right)$, NEC, WEC, and SEC, are violated at the wormhole throat $r=r_{0}$. In other words, we discover that in all plots at the wormhole throat $r=r_{0}$, we have for the energy condition $(\rho^{\rm eff}+p^{\rm eff}_{r})|_{r_{0}=1}<0$, along with the condition $(\rho^{\rm eff}+p^{\rm eff}_{r}+2p^{\rm eff}_{t})|_{r_{0}=1}<0$, by arbitrary small values. Additionally, we see that $\rho^{\rm eff}(r)-|p^{\rm eff}_{r}(r)|<0$ and $\rho^{\rm eff}(r)-|p^{\rm eff}_{t}(r)|<0$ at the WH throat. We will perform the volume integration in Sec.\ref{exotic-matter} to estimate the amount of the total exotic matter that contains in the wormholes.

\section{Amount of exotic matter}
\label{exotic-matter} 
In this section, we consider the “volume integral” which basically provides information about
the “total amount” of averaged null energy condition (ANEC) violating matter in the spacetime. This quantity is related only to $\rho$ and $p_r$, not to the transverse components. It is defined in terms of the following definite integral as \cite{Jusufi:2020rpw}
\begin{eqnarray}
I_V=2 \int^{\infty}_{r_{0}}\big(\rho^{\rm eff}(r)+p^{\rm eff}_{r}(r)\big) dV\,,
\end{eqnarray}
where the volume is given by $dV = r^2drd\Omega$ and $d\Omega$ is solid angle. 
Having introduced a cut-off $r=a$ such that the wormhole extends from $r_0$ to $a$ with $a\geq r_{0} $, we can simply write
\begin{eqnarray}
I_V=8\pi \int^{a}_{r_{0}}\big(\rho^{\rm eff}(r)+p^{\rm eff}_{r}(r)\big)r^{2}dr\,.
\end{eqnarray}
In case of $\Phi=r_{0}/r$, we obtain from the above integral
\begin{eqnarray}
I_V&=&\frac{1}{68040 r^4 r_0^3}\Big(6 r^4 \log (r) \left(5 \pi ^5 \left(3 \sqrt{10}+28\right) \beta -11340 r_0^2 \left(Q^2+3 r_0^2\right)+126 \pi^3 r_0^2\right)\nonumber\\&&+r_0 \Big(-68040 r^2 r_0^2 \left(Q^2 \left(4 r-r_0\right)+2 r r_0^2\right)+5 \pi ^5 \left(3 \sqrt{10}+28\right) \beta  \left(12 r^3+8 r_0^2 r-3 r_0^3\right)\nonumber\\&&+756 \pi^3 r^2 \left(4 r-r_0\right) r_0^2\Big)\Big)\biggr|_{r}^{a}
\end{eqnarray}
Moreover, we can solve for $I_V$ of $\exp(2\Phi(r))=1+\tfrac{\gamma^{2}}{r^{2}}$ to obtain
\begin{eqnarray}
I_V&=& \frac{1}{11340 r_0^3}\Big(\log \left(r^2+1\right) \left(5 \pi ^5 \left(3 \sqrt{10}+28\right) \beta -11340 r_0^2 \left(Q^2+r_0^2\right)+126 \pi ^3 r_0^2\right)\nonumber\\&&\quad\quad\quad\quad+\log (r) \left(-5 \pi ^5 \left(3 \sqrt{10}+28\right) \beta +11340 r_0^2 \left(Q^2+r_0^2\right)-126 \pi ^3 r_0^2\right)\nonumber\\&&\quad\quad\quad\quad+2 r_0^3 \left(5 \pi^5 \left(3 \sqrt{10}+28\right) \beta +11340 \left(Q^2-1\right)-126 \pi ^3\right) \tan ^{-1}(r)\nonumber\\&&\quad\quad\quad\quad+\frac{10 \pi ^5 \left(3 \sqrt{10}+28\right) \beta  r_0^3}{3 r^3}+\frac{10 \pi ^5 \left(3 \sqrt{10}+28\right) \beta  r_0^3}{r}\Big)\biggr|_{r}^{a}\,,
\end{eqnarray}
\begin{figure}[!h]	
\begin{center}		
\includegraphics[width=7.5cm]{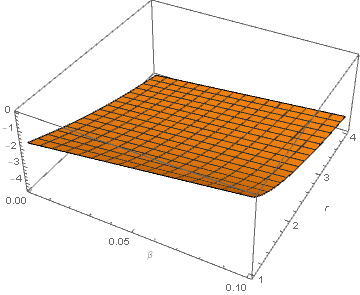}
\includegraphics[width=7.5cm]{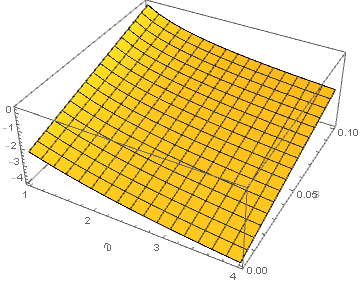}
\caption{We display the variation of $I_{V}$ against $r$ for the case of $\Phi(r)=r_{0}/r$ (left panel)
where we have used $r_{0}=1,\,Q=0.5$ and $\exp(2\Phi(r))=1+\tfrac{\gamma^{2}}{r^{2}}$ (right panel) where we have used $r_{0}=1,\,Q=0.5,\,\gamma=1$. For all plots, we have considered $r\geq r_{0}$  and varied values of $\beta$. \label{Iv}}
\end{center}
\end{figure}
In this specific case, we have used $D_{i}=D_{1}= \frac{5}{42} \left(3 \sqrt{10}+28\right)\pi^2,\,Q= 0.5,\,\gamma=1$ and $r_{0}=1$ for two models.

We observe from Fig.{\ref{Iv}} that the quantity $I_V$ is negative in both cases. This demonstrates the existence
of spacetime geometries containing charged traversable wormholes which are supported by arbitrarily small exotic matter. Such small violations perhaps may be linked to the quantum fluctuations.

\section{Existence of exotic matter and Speed of sound}\label{sec6}
For completeness, one could also add more information of the exoticity.  To ensure the presence of exotic fluid near WH throat, we can use the so-called exoticity parameter \cite{MVisser1995,Capozziello:2011nr} when $\Sigma(r)>0$ implying that exotic matter is present at that point of spacetime. We consider
\begin{eqnarray}
    \Sigma \equiv -\frac{\rho^{\text{eff}} + p_r^{\text{eff}}}{|\rho^{\text{eff}}|}.\label{sig}
\end{eqnarray}
The existence of an exotic matter is required in the range that $\Sigma > 0$. We find from Eq.(\ref{sig}) for $\Phi=r_{0}/r$:
\begin{eqnarray}
\Sigma &=&\frac{r^4}{126 r^7 r_0^3 \left|-90 Q^2+\pi^3+\frac{5\pi^5 \beta \left(3 \sqrt{10}+28\right)}{42 r^2}\right|}\Big(126 r^2 r_0^2 \big(90 \big(Q^2 \big(r^2-4 r_0 r+2 r_0^2\big)\nonumber\\&&\quad\quad\quad\quad\quad\quad\quad+r r_0^2 \big(3 r-2 r_0\big)\big)-\pi ^3 \big(r^2-4 r_0 r+2 r_0^2\big)\big)\nonumber\\&&\quad\quad\quad\quad\quad\quad\quad-5 \left(3 \sqrt{10}+28\right) \pi^5 \beta  \Big(r^4-2 r_0 r^3-4 r_0^3 r+2 r_0^4\Big)\Big).
\end{eqnarray}
Considering points near the WH throat, $r\rightarrow r_{0}$, we find for this case with $\Sigma > 0$
\begin{eqnarray}
\beta >\frac{3780 Q^2-42 \pi^3}{15 \pi^5 \sqrt{10}+140 \pi^5},\,
\end{eqnarray}
which yields $Q>\pi^{3/2}/3\sqrt{10}$. Additionally, for $\exp(2\Phi)=1+\gamma_{2}/r^{2}$, we have
\begin{eqnarray}
\Sigma &=&\frac{1}{126 r^6 \left(r^2+1\right) r_0^3 \left|-90 Q^2+\pi^3+\frac{5\pi^5 \beta \left(3 \sqrt{10}+28\right)}{42 r^2}\right|}\Big(r^4 \big(126 r^3 r_0^2 \Big(-\Big(\pi ^3-90 Q^2\Big)\Big(r^2-1\Big)\nonumber\\&&\quad\quad\quad\quad\quad\quad\quad+2\Big(-90 Q^2+\pi ^3+90\Big) r r_0+90 \left(r^2-1\right) r_0^2\Big)\nonumber\\&&\quad\quad\quad\quad\quad\quad\quad-5 \left(3 \sqrt{10}+28\right) \pi^5 \beta \Big(r^5-r^3-2 \left(2 r^2+1\right) r_0^3\big)\big)\Big).
\end{eqnarray}
Similarly, at points near near WH throat, $r\rightarrow r_{0}$, we find the same condition as that of the previous case
\begin{eqnarray}
\beta>\frac{3780 Q^2 r_0^2-42 \pi ^3 r_0^2}{15 \pi^5 \sqrt{10}+140 \pi ^5}\quad{\rm and}\quad Q>\frac{\pi^{3/2}}{3 \sqrt{10}}\,.
\end{eqnarray}
As mentioned in Ref.\cite{Nandi:2008ij}, the total gravitational energy of a structure composed of normal baryonic matter is negative. Equivalently, we can use total gravitational energy to quantify the fluid given by
\begin{eqnarray}
E_{g}=\frac{1}{2}\int^{r_{1}}_{r_{0}}(1-\sqrt{g_{rr}})\rho^{\rm eff} r^{2}dr+\frac{r_{0}}{2}\,,
\end{eqnarray}
where $g_{rr}$ reads
\begin{eqnarray}
g_{rr}= \Big(1-\frac{b(r)}{r}\Big)^{-1}\,.
\end{eqnarray}
In our models, we can simply show that $E_{g}>0$. As well, a speed of sound (wormhole causality) could be investigated through the equations:
\begin{eqnarray}
    c_{sr}^2 \equiv \frac{d p^{\rm eff}_r}{d \rho^{\rm eff}}, \quad c_{st}^2 \equiv \frac{d p^{\rm eff}_t}{d \rho^{\rm eff}}. \label{csrt}
\end{eqnarray}
Let us start by the case with $\Phi(r)=r_{0}/r$. Substituting $\rho^{\rm eff}$ and $p^{\rm eff}_r$ for this model into Eq.(\ref{csrt}), we find that using $Di=D_{1},\,r=r_{0}=1$, $c_{sr}<1$ for 
\begin{eqnarray}
\beta >\frac{2520 Q^2-28 \pi ^3+1260}{15 \pi ^5 \sqrt{10}+140 \pi ^5},\,
\end{eqnarray}
while $c_{st}=0$ for $r\rightarrow r_{0}=1,\,Q= 0.58$ and $\beta =0.02$. However, for the case with $\exp(2\Phi)=1+\gamma_{2}/r^{2}$, we find that $c_{sr}=0$ for 
\begin{eqnarray}
\beta =7.32\times 10^{-4} \left(90\,Q^2-58.994\right),\,
\end{eqnarray}
while $c_{st}<1$ for $r\rightarrow r_{0}=1,\,Q= 0.58$ and $\beta =0.1$. We note here that the condition of a vanishing sound speed implies that the pressure must be constant. For instance, for model with $\exp(2\Phi)=1+\gamma_{2}/r^{2}$, the radial pressure at the WH throat is constant for $\beta =7.32\times 10^{-4} \left(90\,Q^2-58.994\right)$.

\section{Conclusions}
\label{conclude}
In this work, we investigated new exact and
analytic solutions of the Einstein–Maxwell field equations
describing Casimir wormholes with and without the effect of the GUP corrections. In particular, we considered a specific type of the GUP relation as an example. Along this line, we implemented three specific models of the redshift function along with two different EoS of state given by $p_{r}(r)=\omega_{r}(r)\rho(r)$ along with $p_{t}(r)=\omega_{t}(r)p_{r}(r)$ and obtained the specific relation for the EoS parameter $\omega_{r}(r)$ and $\omega_{t}(r)$, respectively. 

We have obtained a new class of asymptotically flat Casimir wormhole solutions with and without GUP corrections under the effect of electric charge. Furthermore we have checked the null, weak, strong, and dominant energy conditions at the wormhole throat with a radius $r_{0}$, and have demonstrated that in general the classical energy condition are violated by some arbitrary small quantity at the wormhole throat. Importantly, we examined the wormhole geometry with semi-classical corrections via embedding diagrams. We also considered the volume integral quantifier to calculate the amount of the exotic matter near the wormhole throat. For completeness, we have also investigated exotic fluid near WH throat using the so-called exoticity parameter and discussed the speed of sound.

However, to gain more physical interest, the study of anisotropic charged Casimir wormholes with and without GUP corrections may be worth investigating. To do so, we can directly follow relevant machinery proposed for Casimir wormholes found in Ref.\cite{Jusufi:2020rpw}. Additionally, based on the present work, the study of the gravitational lensing effect in the
spacetime of the charged GUP Casimir wormholes with and without GUP correction is another interesting issue to be worth investigating.
\subsection*{Acknowledgments}
DS is financially supported by the Mid-Career Research Grant 2021 from National Research Council of Thailand under a contract No. N41A640145. T. Tangphati was supported by King Mongkut’s University of Technology Thonburi’s Post-doctoral Fellowship. P. Channuie acknowledged the Mid-Career Research Grant 2020 from National Research Council of Thailand (NRCT5-RSA63019-03). This work is partially supported by the National Science, Research and Innovation Fund (SRF) with grant No.R2565B030. 
\subsection*{Data Availability}
Data sharing is not applicable to this article as no data sets were generated or analyzed during the current study.

\end{document}